\documentclass[a4paper,fleqn]{cas-dc}

\usepackage[square, comma, sort&compress, numbers]{natbib}

\def\tsc#1{\csdef{#1}{\textsc{\lowercase{#1}}\xspace}}
\tsc{WGM}
\tsc{QE}
\tsc{EP}
\tsc{PMS}
\tsc{BEC}
\tsc{DE}
\usepackage{bm}
\usepackage{amsmath}

\begin{document}
	\let\WriteBookmarks\relax
	\def\floatpagepagefraction{1}
	\def\textpagefraction{.001}
	
	 \shorttitle{Engineering Applications of Artificial Intelligence}
	
	\shortauthors{Q. Huang,L. Jia et~al.}
	
	\title [mode = title]{Extraction of Vascular Wall in Carotid Ultrasound via a Novel Boundary-Delineation Network}                      

	\author[1,2,3]{Qinghua Huang}[]
	
	\cormark[1]
	
	\ead{enicarwhw@qq.com}
	
	\author[1,2]{Lizhi Jia}[]
	\author[4]{Guanqing Ren}[]
	\author[4]{Xiaoyi Wang}[]
	\cormark[1]
	\ead{Gm@delicasz.com}
	\author[5]{Chunying Liu}[]
	\cormark[1]
	\ead{375529287@qq.com}
	

    \affiliation[1]{organization={School of Mechanical Engineering, Northwestern Polytechnical University},	 
		state={Xi'an},
		postcode={710072},
		country={China}}
   		\affiliation[2]{organization={School of Artificial Intelligence, Optics and Electronics (iOPEN), Northwestern Polytechnical University},	 
   		state={Xi'an},
   		postcode={710072},
   		country={China}}
  
   \affiliation[3]{organization={School of Electronic Information, Guangxi University for Nationalities},
   	state={Nanning},
   	postcode={530006}, 
   	country={China}}

   	\affiliation[4]{organization={Shenzhen Delica Medical Equipement Company Limited},
   		state={Shenzhen},         
   		postcode={518132},       
   		country={China}}
    
   \affiliation[5]{organization={Hospital of Northwestern Polytechnical University},	 
   	state={Xi'an},               
   	postcode={710072},
   	country={China}}

	\cortext[cor1]{Corresponding author}

	\begin{abstract}
		Ultrasound imaging plays an important role in the diagnosis of vascular lesions. Accurate segmentation of the vascular wall is important for the prevention, diagnosis and treatment of vascular diseases. However, existing methods have inaccurate localization of the vascular wall boundary. Segmentation errors occur in discontinuous vascular wall boundaries and dark boundaries. To overcome these problems, we propose a new boundary-delineation network (BDNet). We use the boundary refinement module to re-delineate the boundary of the vascular wall to obtain the correct boundary location. We designed the feature extraction module to extract and fuse multi-scale features and different receptive field features to solve the problem of dark boundaries and discontinuous boundaries. We use a new loss function to optimize the model. The interference of class imbalance on model optimization is prevented to obtain finer and smoother boundaries. Finally, to facilitate clinical applications, we design the model to be lightweight. Experimental results show that our model achieves the best segmentation results and significantly reduces memory consumption compared to existing models for the dataset.
	\end{abstract}
	
	

	\begin{highlights}
		\item Propose the boundary-delineation network (BDNet) for vascular wall segmentation of carotid ultrasound images.
		\item Using boundary refinement module to precisely locate the boundary points of the vascular wall to prevent boundary localization errors.
		\item Design feature extraction module to enhance the extraction and fusion of low-quality image features.
		\item  Propose loss function to focus the optimization of the model on vascular wall segmentation.
		\item  To facilitate clinical applications, the BDNet is designed to be lightweight.
	\end{highlights}
	
	\begin{keywords}
		Boundary delineation  \sep Ultrasound image segmentation \sep Vascular wall \sep Low quality images \sep Lightweight  
	\end{keywords}

	\maketitle
	
	\section{Introduction}
	
	In 2018, 17.9 million people died from cardiovascular disease (CVD) worldwide, 16.2\% higher than reported in 2006 \cite{benjamin2019heart}. Cardiovascular disease has become the leading cause of death and disability worldwide and will continue to grow. Especially in low-and middle-income countries, the incidence of CVD is increasing every year \cite{bansilal2015global}. There is an independent association between vascular and cardiovascular disease risk \cite{sedaghat2018common}. Vascular examinations are a major part of cardiovascular disease screening. Therefore, the work on the analysis of blood vessels is particularly important. The prevention of cardiovascular diseases is urgent. For the diagnosis of blood vessels, the vascular wall can be a good reflection of the structure of the vessels. The vascular wall structure is analyzed to arrive ﬁnal diagnosis.

	The main clinical methods of vascular examination are diagnostic pathology, ultrasound, CT and MRI \cite{liu2019size}. Ultrasound imaging techniques are widely used because of their safety, low cost and non-invasive nature \cite{sanches2012ultrasound}. To perform ultrasound examinations of blood vessels, two physicians are required to collaborate. One is responsible for scanning the patient and the other is responsible for labeling the images for diagnosis. Errors can occur due to the quality of imaging and the subjective awareness of the physician. In order to reduce the burden on the physician, speed up the diagnosis and improve diagnostic accuracy, computer-aided systems (CAD) are beginning to appear on the horizon.

	In the beginning, researchers mostly used traditional operators for the identification and segmentation of vascular structures. The commonly used traditional algorithms are gradient calculation, edge tracking, dynamic programming, active contour, hough transform, snakes, etc \cite{molinari2010state}. For example, MA Gutierrez et al. \cite{gutierrez2002automatic} proposed an active contour improvement technique based on multi-resolution analysis. The model uses filters of different scales to extract the boundaries by derivation. MC Bastida-Jumilla et al. \cite{bastida2013segmentation} used the sobel operator to detect the horizontal edge of the artery. The hough transform detects the main direction of the artery. Finally, active contours segment the carotid artery walls. Filippo Molinari et al. \cite{molinari2010integrated} treated the artery as consisting of a lumen with a low-intensity region and two bright borders. Many seed points were randomly generated in the image. By connecting these seed points, the vascular wall structure was formed. A discriminator was used to reject the unqualified boundaries and retain the true vascular wall boundaries.

	In recent years, with the rapid development of deep learning in the field of medical image processing, researchers have started to focus their research on deep learning algorithms. Compared with traditional methods, deep learning algorithms have better robustness. Several researchers have started to use deep learning methods to solve the vascular structure segmentation problem. Researchers started to use U-Net and its variants to segment the structure of blood vessels \cite{xie2019vessel,szarski2021improved,yang2018ivus,yang2019robust,azzopardi2020bimodal}. As the research progresses, various models were proposed \cite{lian2021april}.

	For example, Carl Azzopardi et al. \cite{azzopardi2017automatic} used an encoder-decoder structure to complete the segmentation of the vascular by two downsamples and two upsamples. Later, U-Net \cite{azzopardi2020bimodal} was proposed to segment the cross-section of the lumen of the vessel derive the vascular wall structure. Xie et al. \cite{xie2019vessel} segmented the longitudinal section of the lumen of a blood vessel by using a two-path U-Net model. The original image and the image after data enhancement were used as two inputs, and then the features extracted from both paths were predicted using a decoder. Yang et al. \cite{yang2019robust} used U-Net as the backbone. A special design was performed in skip connection. A parallel structure is used to refine the downsampled features. The two branches are the main branch and the detailed branch, and the main branch uses two layers of convolutional layers with 5×5 kernel size. The detail branch uses a convolutional layer of 3×3 kernel size and a convolutional layer of 1×1 kernel size. The final prediction is performed using 5×5 convolution for the prediction of the results and expanding the receptive fields. Martin Szarski et al. \cite{szarski2021improved} tried to add coordinate information inside the U-Net model. It is hoped that the spatial location dependencies of the features can be identified. The authors added CoordConv after the first downsampling of U-Net and added spatial coordinate information through CoordConv. The other parts remain unchanged. Lian et al. \cite{lian2021april} used four SE-ResNeXt modules, an average pooling layer and two fully connected layers to construct the feature encoder. The decoder combined with the DQN network is used to generate the mask and five key points. Five key points are used to constrain the shape of the mask.

	We found that previous researchers had taken a variety of approaches to improve the accuracy of vascular wall segmentation. Although those methods mentioned above have achieved good results in vascular structure segmentation. But there are still some problems with the present methods. First, most of the existing methods down-sample and up-sample the image by convolution to obtain the segmentation result of the vascular wall. The upsampling process can lead to the blurring of vascular wall boundary information, which will result in inaccurate boundary locations \cite{kirillov2020pointrend}. The second one is that ultrasound devices do not always output high-resolution, good-quality images. Due to the noise of the device or the distortion of the images, there are many images with bad quality in real situations. For example, there are images with dark boundaries, discontinuity boundaries and other problems. We should design the model with these problems in mind and try to avoid being affected by them. The third problem is that the ultrasound image vascular wall segmentation task is a class imbalance between the vascular wall and the background. If no approach is taken to deal with it, the segmentation accuracy of the model cannot be well optimized when training the model. The last one is that the present segmentation algorithm has a large number of backbone parameters, which require high hardware for the device. It is not conducive to clinical application and popularity.

	Based on the above problems, a boundary-delineation network (BDNet) is designed to solve the vascular wall boundary delineation problem. The precise positioning of boundary points and extraction of boundary information is achieved by boundary refinement to prevent the offset of the boundary. The problem of dark boundaries and discontinuity boundaries in lower-quality images is well solved by extracting multi-scale features and fusing different receptive field features. A combination of the lovász-softmax loss function and cross-entropy loss function is used to avoid the class imbalance problem of the vascular wall segmentation task. The boundary refinement is optimized using point cross-entropy loss. In order to reduce the hardware requirements and the number of model parameters, we have designed the model to be lightweight. In summary, the main contributions of our paper are as follows.

	\begin{enumerate}
		\itemsep=0pt
		\item Adopt the boundary refinement to accurately locate and extract features from the boundary points of ultrasound image vascular wall, and prevent the problem of inaccurate boundary positioning during the upsampling process.
		\item Introduce a multi-scale fusion mechanism and PSP module to fuse multi-scale and different receptive field features to solve the problem of incorrect segmentation in low-quality images with dark boundaries and discontinuity boundaries.
		\item Using a new loss combination to solve the problem of class imbalance between the vascular wall and the background, putting the center of model optimization on the segmentation of the vascular wall.
		\item Lightweight optimization of the model to significantly reduce the number of parameters while maintaining the model accuracy.

	\end{enumerate}  
	
	The rest of the paper is structured as follows. Section \ref{Section II} describes the methodology of the model. Section \ref{Section III} describes the details of the experiments. Section \ref{Section IV} provides a discussion. Section \ref{Section V} draws the conclusion of the paper.
	\section{Methods}
	\label{Section II}
	\begin{figure*}
		\begin{center}		
			\includegraphics[width=\textwidth]{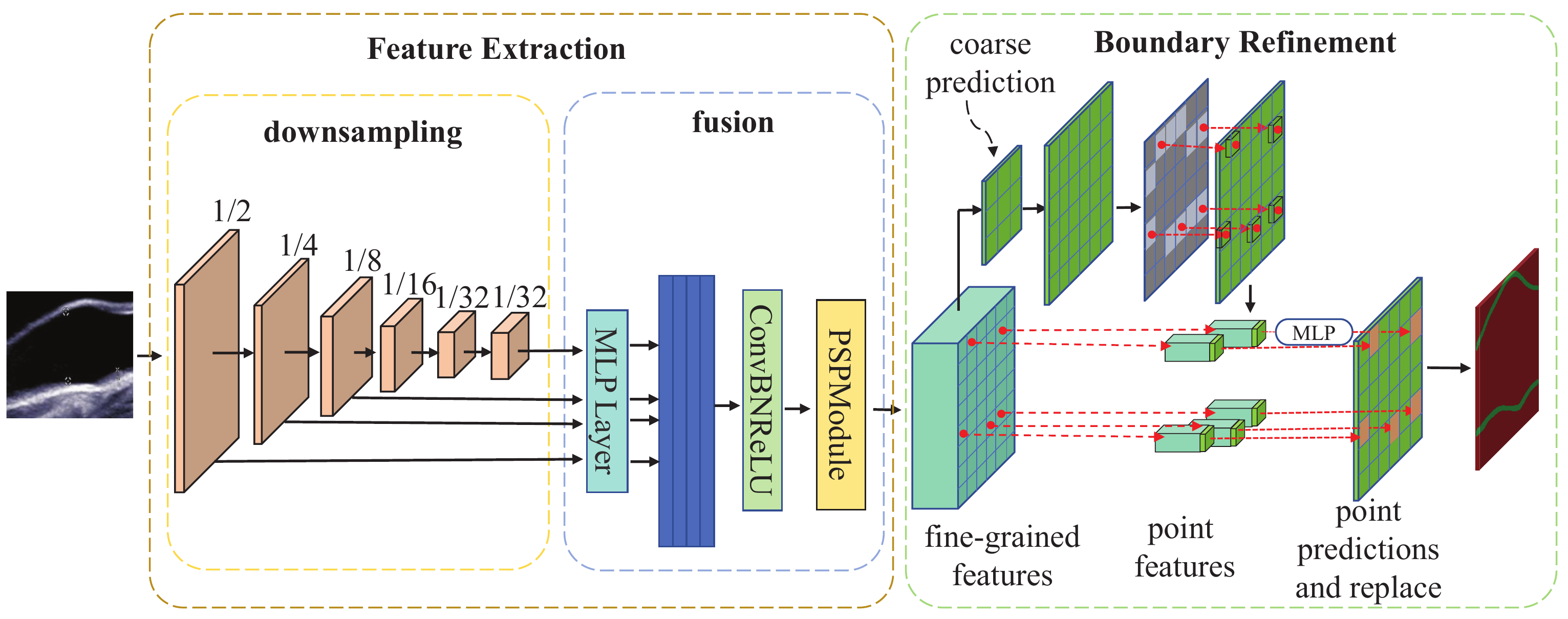}		
			\caption{Overview of the BDNet structure. There are mainly two modules. 1) The feature extraction module extracts image features  and fuses multi-scale features and different receptive field features to output fine-grained features. 2) Boundary refinement module is used to extract the boundary points, re-predict the boundary points and replace the previous coarse prediction results.}
			\vspace{-2em}
			\label{fig1}
		\end{center}
		
	\end{figure*}

	In this section, we describe the structure of the BDNet in detail. BDNet consists of four modules, which are the feature extraction module, the boundary refinement module, the loss function module and the lightweight module, the whole model is shown in Fig. \ref{fig1}. The image is first input into the feature extraction module. The feature extraction module is designed for feature downsampling and feature fusion. In the feature downsampling process, the multi-scale features are saved. In the process of feature fusion, the multi-scale features are fused, and the different receptive field features are generated and fused. Then, the output of the feature extraction module is input into the boundary refinement module for boundary position correction. The locations of suspected boundary points are extracted. The fine-grained features of these points are re-predicted using MLP. The results of these points in the coarse prediction are replaced and the final prediction results are output. After that, the prediction results are computed by our new loss function to better optimize the model and prevent the effect of class imbalance. Finally, we developed a lightweight downsampling module for our proposed BDNet to speed up the training and testing process, making it suitable for clinical applications and generalization.
	
	\subsection{Feature extraction}
	In clinical practice, there are a large number of ultrasound vascular images that are of low quality. Common problems include discontinuous boundaries and dark boundaries.
	
	Dark boundaries refer to the fact that the boundary of the vascular wall in a certain region has a small gray value compared to other regions and is not obvious, which is easily ignored in the downsampling process. Dark boundaries can lead to broken segmentation results when dark boundaries are not detected. The lower-level features contain more detailed information. We can use the low-level features to prevent the problem of losing information on dark boundaries. Therefore, we use the combination of low-level features and high-level features for multi-scale features to predict the results.
	
	Discontinuous boundaries refer to the appearance of intermittent connections in the vascular wall in the ultrasound image. Discontinuous boundaries can lead to intermittent segmentation results that are not conducive to application. In ultrasound images, blood vessels are generally throughout the image. The vascular wall is regular. In the case of discontinuities at the boundaries of the vessels, sometimes it is necessary to use information from other regions for prediction. If only downsampling is used, only information from adjacent areas can be extracted \cite{kirillov2020pointrend}. If the discontinuity boundary is long, there is no boundary around it, this will lead to prediction failure. We can introduce more contextual information and establish spatial dependencies to prevent prediction failure when the discontinuity boundaries are long. Therefore we use the PSP module, using four receptive fields of view to introduce more contextual information. The discontinuity boundaries are correct from global and local perspectives.
	
	    \begin{figure}
		\begin{center}
			\includegraphics[width=\columnwidth]{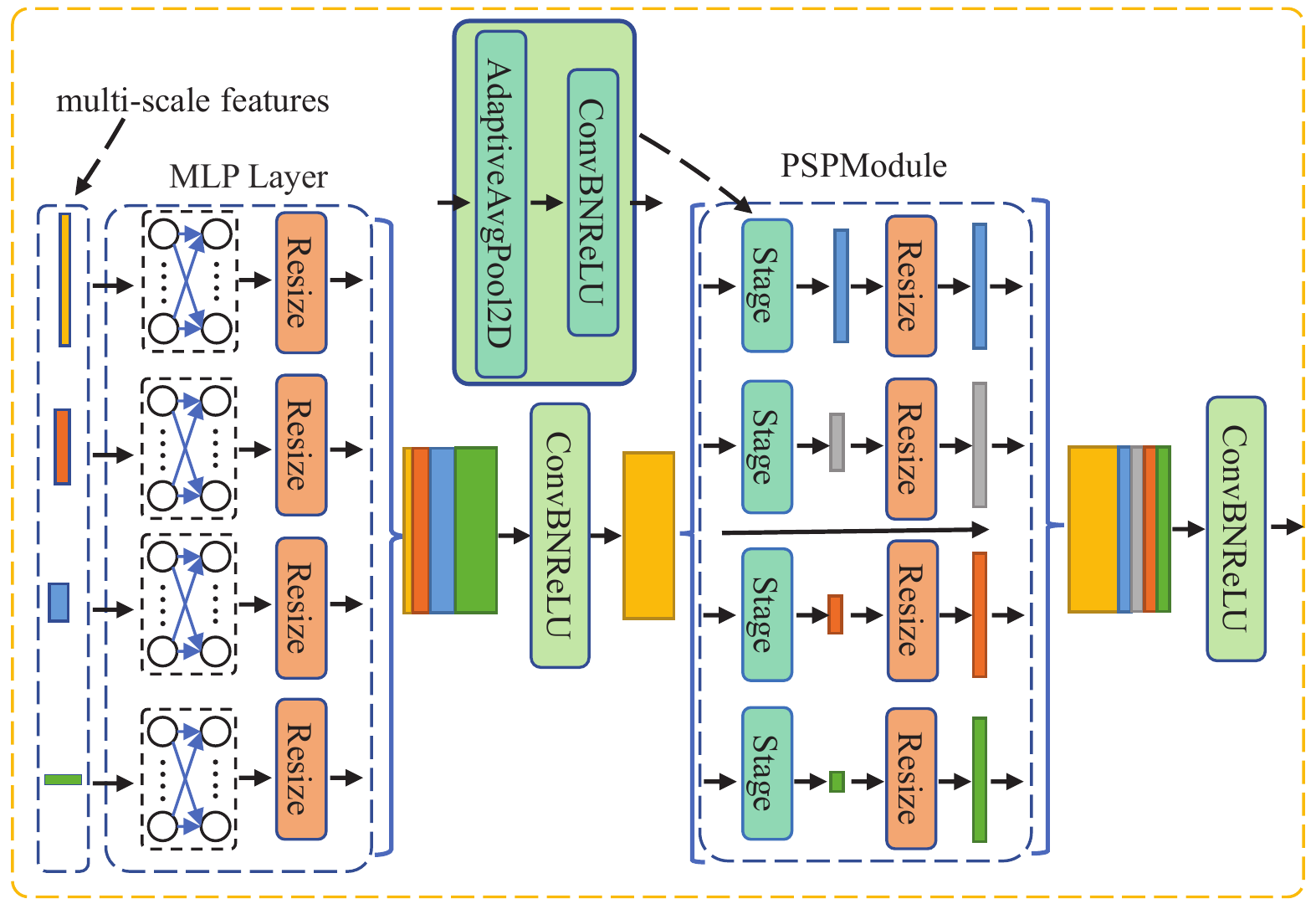}		
			\caption{The architecture of the fusion module. The left of the figure performs the fusion of multi-scale features. The right is performing the fusion of different receptive field features.}
			\label{fig3}
		\end{center}	
	\end{figure}
	
	For ultrasound images, boundary discontinuities and dark boundaries are inevitable. However, we can take methods in the feature extraction module to avoid the effect of these situations on the segmentation results. The feature extraction module consists of the downsampling module and the fusion module. 
	
	The purpose of the downsampling module is to extract the high-level features of the image and retain the multi-scale features. We perform five convolution calculations with a stride of 2 on the input image features, as shown in Fig. \ref{fig1}. The features are changed to 1/32 of the original image. After that, we perform another convolution calculation with a stride of 1 to embed the features. During downsampling, we preserve features of 1/2, 1/4, and 1/8 size of the original image and the embedded features.
    
    The features of 1/2, 1/4, 1/8 and 1/32 saved in the downsampling module will be fused in the fusion module, as shown in Fig. \ref{fig3}. Before multi-scale fusion, we use the MLP layer to resize the image features of different scales to the same size for easy fusing afterward. The MLP also has some role in fusing spatial and channel features \cite{tolstikhin2021mlp}, it is used as a small auxiliary module. After that, we aggregate multi-scale features by concatenating. The multi-scale feature fusion is done by convolution, batch normalization and ReLU activation functions.

	We need to get information about different receptive fields. We use the PSP module \cite{zhao2017pyramid} to implement. The local information and global information are fused to introduce more contextual information. We set up four different fields of view, 1×1, 2×2, 3×3 and 6×6, as shown in Fig. \ref{fig3}. In each stage, we perform an adaptive averaging pooling of the features to the size we want. To facilitate feature fusion, we resize the features to 1/8 of the original image. After that, the features are extracted by convolution, batch normalize and ReLU activation functions. The global information is fused with the local information. Finally, we fuse the pre-processed features with the post-processed features to avoid information loss. 
	
	\subsection{Boundary refinement}
	The general strategy used for ultrasound image segmentation is to downsample to obtain high-density features and then upsample to the original image size. The process of upsampling linear interpolation blurs the boundary information and leads to the smoothing of boundary features. This phenomenon causes the area of the predicted mask to be larger than the label. For example, the lumen of a blood vessel after upsampling back to the original image size will have the lumen contour offset outward than the true contour. The same is true for the contours of the vascular wall tissue. Therefore, we refer to the wrong contour when performing softmax resulting in inaccurate localization of the boundaries. To prevent the boundary offset, we introduce the boundary refinement module to help solve this problem. We use the PointHead \cite{kirillov2020pointrend} to implement the boundary refinement. With boundary refinement, we can achieve precise positioning of boundary points and prevent boundary positioning errors.
	\begin{figure}
		\begin{center}
			\includegraphics[width=\columnwidth]{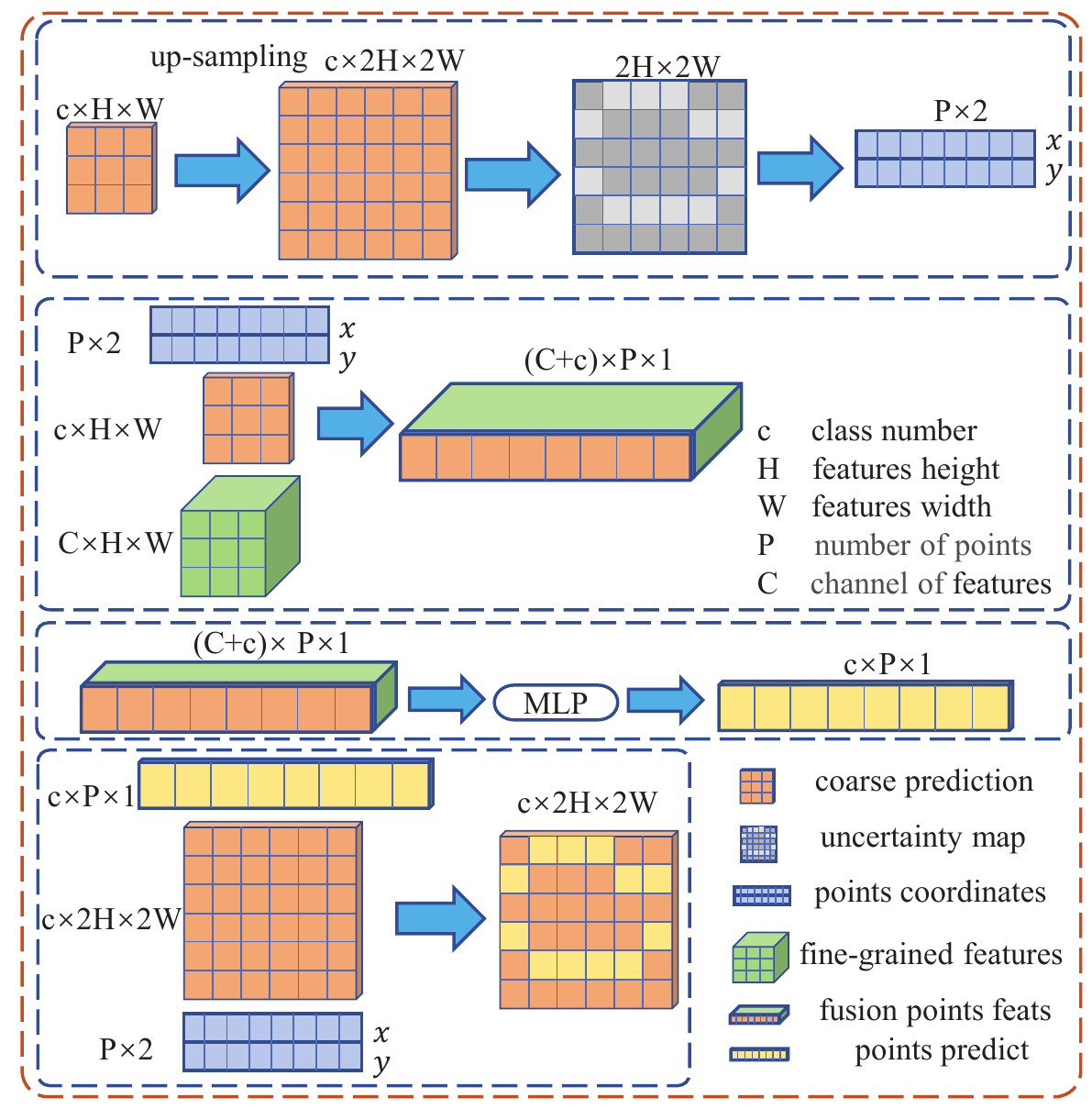}		
			\caption{Detailed implementation of the boundary refinement module. The process is divided into four steps: finding uncertain points, extraction of features corresponding to the points, re-prediction of the points, and replacement of the points.}
			\label{fig4}
		\end{center}	
	\end{figure}
	
	The boundary refinement module is divided into four steps, as shown in Fig. \ref{fig4}. First, upsampling is performed on the coarse prediction results. Based on the result after upsampling, the uncertainty map of the classification result is calculated. The first $P$ uncertain points are selected according to the uncertainty map and their coordinates are determined. For each pixel, the size of the model output result is $c$×1. $c$ is the number of classes. The output of each pixel point is denoted by $F_{point}$. The formula for calculating the uncertainty of each pixel is as follows.

    \begin{equation}
	\label{eq2}
	S_{top1},S_{top2} = Topk\left( F_{point},k = 2 \right)
    \end{equation}
    \begin{equation}
	\label{eq3}
	     U = S_{top1}{- S}_{top2}
    \end{equation}	

	$Topk$ is a function that calculates the scores of the top $k$ classes with the highest scores for each pixel. $S_{top1}$ denotes the score of the first ranked class. $S_{top2}$ denotes the score of the second ranked class. $U$ denotes the uncertainty of the pixel point.

	After determining these points, we obtain the corresponding coordinates. The coordinates are normalized. If the image feature size changes, it does not affect the determination of the position. Based on the coordinates, we find the features at the corresponding positions from coarse prediction and fine-grained features. The two features are concatenated according to the channel direction. Then, the features are re-predicted for each pixel after fusion using an MLP network. Finally, the result of the coarse prediction corresponding to the position is replaced with the new prediction result according to the coordinates.

	The boundary is the dividing line between the two class areas. Uncertain points mean that the scores of the different classes are similar at this location. This may be caused by the overlapping of regions due to the expansion of different classes of regions during the upsampling process. Therefore, it indicates that
	uncertain points may be boundary points. With the boundary refinement module, we correct the incorrect results of upsampling by re-predicting the labels of the boundary points based on fine-grained features without damaging the detailed information of the boundary. We can prevent the boundary position from offset due to upsampling. Thus, an accurate boundary position is obtained. In addition, the prediction of carefully selected points is computationally an order of magnitude less compared to direct computation. This can speed up the segmentation of the model.
	
	\subsection{Loss}
	During the training of the model, the model outputs an ensemble of points and a mask. The training of the set of points is handled by the point cross entropy loss ($L_{PCE}$). The training of the mask is handled by cross entropy loss ($L_{CE}$) and lovász-softmax loss ($L_{LS}$) \cite{berman2018lovasz}.

	$L_{PCE}$ is a loss function designed for the boundary refinement module. The model uses this loss function to improve the performance of boundary point re-prediction. The essence of point cross entropy loss is cross entropy loss, but it is calculated for the set of points. The formula of $L_{PCE}$ is shown below.
	
\begin{equation}
	\label{eq4}
	CE = \frac{1}{N}{\sum\limits_{i = 1}^{N}\left( - {\sum\limits_{j = 1}^{C}y_{i,j}}logp_{i,j} \right)}
\end{equation}
\begin{equation}
	\label{eq5}
	L_{PCE} = CE\left( {points}_{pre},{points}_{label} \right)
\end{equation}

	$CE$ denotes cross entropy. ${points}_{pre}$ denotes the prediction of uncertainty points generated by the boundary refinement module. Its size is $N$×$C$×$Nums$. $N$ is the batch size. $Nums$ is the number of uncertainty points. $C$ denotes the classes of the segmentation. ${points}_{label}$ indicates the true value of these uncertainty points. Its size is $N$×$Nums$. $y_{i,j}$ is the true label of sample $i$ on class $j$. $ p_{i,j}$ is the predicted probability of sample $i$ on class $j$.

	$L_{CE}$ is the main loss of loss function, which is used to calculate the loss of the mask. ${mask}_{pre}$ is the prediction mask. Its size is $N$×$C$×$H$×$W$. $H, W$ denote the height and width of the image respectively. ${mask}_{label}$ is the true value of the mask. Its size is $N$×$H$×$W$.
\begin{equation}
	\label{eq6}
	L_{CE} = CE\left( {mask}_{pre},{mask}_{label} \right)
\end{equation}

	Vascular wall segmentation is a class unbalanced segmentation task. The background occupies a large proportion and the region of the vascular wall is small. When evaluating the model performance, the segmentation accuracy of the background pulls up the segmentation accuracy of the model, which is bad for the optimization of our model. When optimizing the model parameters, it is not possible to optimize the vascular wall segmentation accuracy. We introduced the lovász-softmax loss to solve the class imbalance problem \cite{wang2020human} and optimize the model better. Lovász-softmax loss calculates the mIoU of the model segmentation results. The mIoU can better reflect the segmentation accuracy of the imbalance class than $L_{CE}$. Thus, the segmentation accuracy of the vascular wall is improved and the segmentation results are smoother \cite{xuhong2021land}. The formula of $L_{LS}$ is shown below.
\begin{equation}
	\label{eq7}
	m_{i}(c) = \left\{ \begin{matrix}
		{1 - f_{i}(c)~~if~c = y_{i}^{*}} \\
		{~~~~~f_{i}(c)~~~~~otherwise~} \\
	\end{matrix} \right.
\end{equation}
\begin{equation}
	\label{eq8}
	LS = \frac{1}{|C|}{\sum\limits_{c \in C}{\overset{-}{\mathrm{\Delta}_{j_{c}}}\left( m(c) \right)}}
\end{equation}
\begin{equation}
	\label{eq9}
	L_{LS} = ~LS\left( {mask}_{pre},{mask}_{label} \right)
\end{equation}

	${\overset{-}{\mathrm{\Delta}_{j_{c}}}}$is the lovász extension of IoU  \cite{berman2018lovasz}. $C$ indicates the class of segmentation. $y_i^\ast$ indicates the ground truth class of pixel $i$, $f_i\left(c\right)$ indicates the class probabilities of pixel $i$.
	
	Our loss function is as follows:
\begin{equation}
	\label{eq10}
	Loss = ~L_{CE} + {(1 - \alpha)L}_{LS} + \alpha L_{PCE}
\end{equation}

	The optimization of the model is focused on the segmentation of the vascular wall through the combination of our loss functions. In the training process, we use two stages of training. The loss function has two combinations. $\alpha$ takes 0 or 1. We use $L_{CE}$ and $L_{PCE}$ for training in the early stage of training to capture the image boundary details better and to speed up the convergence. We take $\alpha$ to 1. At the later stage of training, we take $\alpha$ to 0 in order to make the segmentation more accurate and the boundary smoother. We use $L_{CE}$ and $L_{LS}$ for the finetune.
	
	\subsection{Lightweight}
		\begin{figure*}
		\begin{center}
			\includegraphics[width=\textwidth]{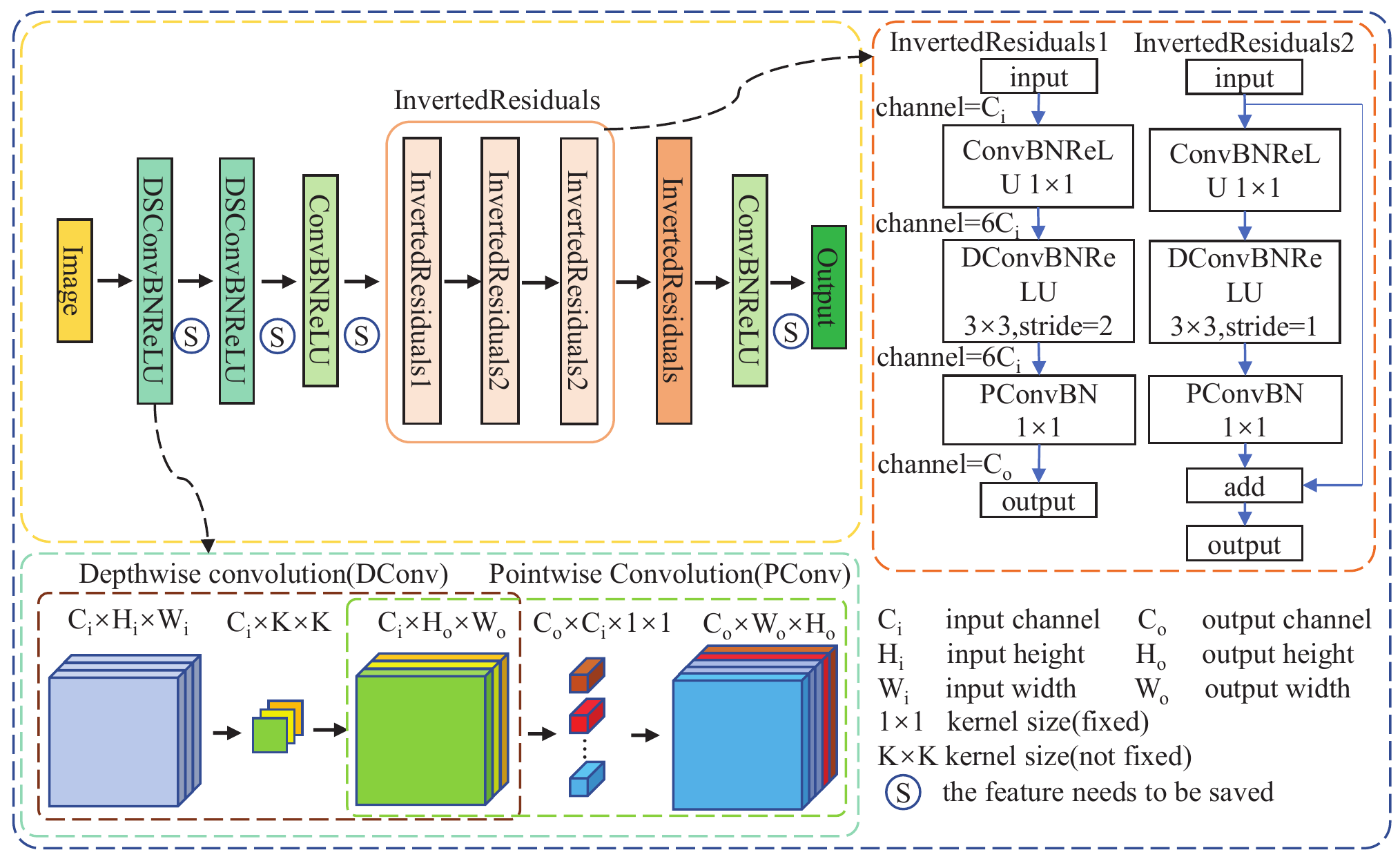}		
			\caption{Detailed implementation of the lightweight downsampling module. The light blue dashed box shows the implementation principle of depthwise separable convolution. The brown dashed box is the implementation of the inverted residuals.}
			\label{fig2}
		\end{center}	
	\end{figure*}
		
	A large number of parameters and complex computations severely limit the use of deep learning products in daily life. In order to make vascular wall segmentation faster and take up less computational resources than existing methods to speed up the use of artificial intelligence products in clinical settings. We need to design the model to be lightweight. Starting from this direction, we designed a lightweight downsampling module to extract features from images by depthwise separable convolution \cite{howard2017mobilenets} and inverted residuals \cite{sandler2018mobilenetv2}. The flow of the whole module is shown in Fig. \ref{fig2}. Save in Fig. \ref{fig2} means that the features at this location need to be saved. Fixed means that the size of the convolution kernel must be 1×1. Not fixed means that the size of the convolution kernel can be determined by yourself according to your model. The purpose of the whole module is to reduce the number of parameters.
	
	The depthwise separable convolution divides the convolution process into depthwise convolutions and pointwise convolutions. Depthwise convolution is a channel corresponding to only one convolution kernel. Pointwise convolution uses a 1×1 convolution kernel to extend the depth, as shown in Fig. \ref{fig2}. Depthwise separable convolution significantly reduces the parameters of the convolution and the computational effort of the model. Fewer parameters are important for the application of the system. The equation (\ref{eq1}) shows the ratio of the number of parameters between traditional convolution and depthwise separable convolution. $P_{DSConv}$ is the number of parameters of depthwise separable convolution. $P_{Conv}$ is the number of parameters of traditional convolution. $C_i$ denotes the number of input channels. $C_o$ is the number of output channels. $K$ is the convolution kernel size.
	\begin{equation}
		\label{eq1}
		\frac{P_{DSConv}}{P_{Conv}} = \frac{C_{i} \bullet K \bullet K + C_{o} \bullet C_{i}}{C_{o} \bullet {C}_{i} \bullet K \bullet K} = \frac{1}{C_{o}} + \frac{1}{K^{2}}
	\end{equation}
	
	Lightweight networks generally have a smaller number of channels in order to reduce the parameters and speed up the computation. However, the reduction in the number of channels can lead to a model that does not perform well for image information extraction \cite{sandler2018mobilenetv2}. Therefore, the inverted residuals have two roles. One is to extract advanced image features, and the other is to avoid the problem of weakened image information extraction caused by a smaller number of channels. The implementation details of the inverted residuals are shown in Fig. \ref{fig2}. In order to reduce the number of parameters, the depthwise separable convolution is used. Inverted residuals mainly expand the number of channels and compress the number of channels for image features. By increasing the number of channels, the model can extract features well.

    \section{Experiments}
    \label{Section III}
    \subsection{Dataset}
    This study is approved by the Ethics Committee of Medical and Experimental Animals, Northwestern Polytechnical University, Xi’an, China (Protocol no. 202002010). We obtained longitudinal carotid ultrasound images of 657 patients from our partner medical institutions. Each image was labeled by the physician with the region of interest. We cropped the ROI from the original images. The part of the vascular wall in each ROI image was labeled by the physician. These annotations were used to generate mask images as labels. 1548 vascular images were finally obtained. This dataset is challenging. The data comes from different hospitals and different brands of ultrasound equipment. The quality of each image was good or bad. The region of interest includes several different sites, and the shape of the blood vessels is different in each site.

   \subsection{Evaluation metrics}
   To better evaluate the performance of our model, we use five metrics to evaluate the segmentation performance of our model. These include Dice, mIoU, boundary IoU (BIoU), number of params (Params), and floating point operations (FLOPs). mIOU is the average of the intersection of the true and predicted values of the different classes, and takes the value [0-1]. mIOU is larger to indicate better model performance. Dice can calculate the similarity between two samples. The value is [0-1]. the larger Dice, the better performance. BIoU is a dedicated metric to measure the goodness of the boundary segmentation model \cite{cheng2021boundary}. FLOPs can be used to measure the computational complexity of the algorithm. Params directly determine the size of the model. The formulas of the metrics are as follows.
\begin{equation}
	\label{eq11}
	mIOU = \frac{1}{k + 1}{\sum\limits_{i = 0}^{k}\frac{TP}{PN + FP + TP}}
\end{equation}
\begin{equation}
	\label{eq12}
	Dice = ~\frac{2 \bullet TP}{2 \bullet TP + PN + FP}
\end{equation}
   
   $TP$ denotes positive samples with correct prediction, $FP$ denotes positive samples with incorrect prediction, $FN$ denotes negative samples with incorrect prediction, and $TN$ denotes negative samples with correct prediction.
\begin{equation}
	\label{eq13}
	BIoU = \frac{\left| {\left( {G_{d} \cap G} \right) \cap \left( {P_{d} \cap P} \right)} \right|}{\left| {\left( {G_{d} \cap G} \right) \cup \left( {P_{d} \cap P} \right)} \right|}
\end{equation}
  
   $G$ denotes the ground truth binary mask. $P$ denotes the prediction binary mask. $G_d$, $P_d$ denotes the set of pixels in the boundary region of the binary mask.

   \subsection{Experimental details}
   Our experiments relied on a hardware environment with the server equipped with the Quadro RTX 8000 GPU. The operating system was Ubuntu. We used Paddle as our deep learning framework. Our dataset was divided into a training set and a test set with a ratio of 4:1. We trained and evaluated the model with five-fold cross-validation. During training, the images were randomly cropped, randomly flipped, randomly scaled and normalized. Only normalization was used in the evaluation process. Our experiments were divided into two training phases. In the first stage, the $\alpha$ of the loss function was set to 1 and trained for 200 epochs. In the second stage, the $\alpha$ of the loss function was set to 0 and trained for 100 epochs.
   
   \subsection{Comparative Study}
   In order to be able to better evaluate the performance of our model, we compared the current dominant general-purpose segmentation models. In addition, we compared previous models that have done the same work.
   
   \subsubsection{Comparison with general models}
   \begin{figure*}
   	\begin{center}				
   		\includegraphics[width=\textwidth]{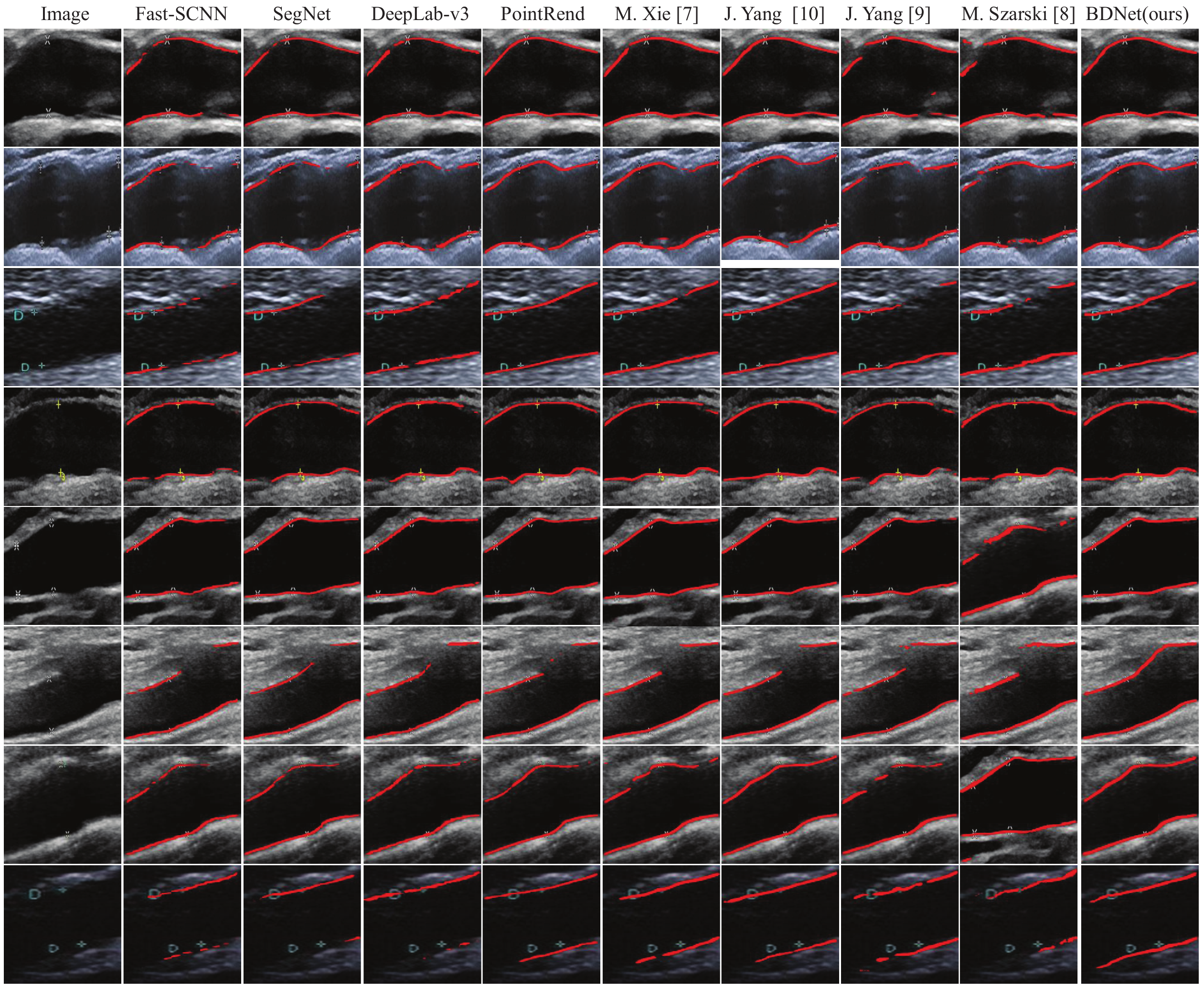}
   		\vspace{-2em}
   		\caption{Visualization of segmentation results for different models.}
   		\vspace{-2em}
   		\label{fig5}
   	\end{center}	
   \end{figure*}
 
   We compared our results with existing generic methods. BiSeNetV2 \cite{yu2021bisenet} and Fast-SCNN \cite{poudel2019fast} are the latest lightweight segmentation methods. U-Net \cite{ronneberger2015u}, SegNet \cite{badrinarayanan2017segnet}, DeepLab-v3 \cite{chen2017rethinking}, PSPNet \cite{zhao2017pyramid} and PointRend \cite{kirillov2020pointrend} are the more popular and used methods nowadays.
   \begin{table}[]
   	\caption{Quantitative comparison of different generic models, using evaluation metrics for both the model itself and the segmentation results obtained from the model.}
   	\label{tab1}
   	\begin{center}
   		\resizebox{\columnwidth}{!}{
   			\begin{tabular}{cccccc}
   				\toprule  
   				Model               & Dice & mIoU & BIoU & Params(M) & FLOPs(G) \\ \midrule
   				BiSeNetV2 \cite{yu2021bisenet}  & 58.2 & 69.9 & 42.4 & 2.2       & 2.9      \\
   				Fast-SCNN \cite{poudel2019fast}  & 60.3 & 70.9 & 44.5 & 1.4       & 0.4      \\ \midrule
   				U-Net \cite{ronneberger2015u}      & 60.5 & 70.8 & 44.7 & 12.8      & 45.2     \\
   				SegNet \cite{badrinarayanan2017segnet}     & 60.9 & 71.0 & 45.1 & 28.2      & 61.9     \\
   				DeepLab-v3 \cite{chen2017rethinking} & 62.1 & 71.5 & 46.2 & 37.3      & 59.2     \\
   				PSPNet \cite{zhao2017pyramid}     & 63.1 & 71.8 & 47.2 & 64.8      & 96.6     \\
   				PointRend \cite{kirillov2020pointrend}  & 63.3 & 71.9 & 47.6 & 26.9      & 68.1     \\ \midrule
   				BDNet (ours)                & 65.6 & 73.2 & 50.0 & 1.3       & 1.9      \\ \bottomrule
   			\end{tabular}
   		}
   	\end{center}
   \end{table}
  
   As shown in Table \ref{tab1}, our model produced 65.6 Dice using only 1.3M Params and 1.9G FLOPs, higher than the results of other methods. Compared with Fast-SCNN, our model had a slightly higher number of Params and FLOPs, but our Dice segmentation result was 5.3 higher compared with Fast-SCNN. Compared with other models, our model was superior in both the number of parameters and segmentation accuracy. Fig. \ref{fig5} shows the segmentation results of our model and other models. We can see that our model can solve the problem of dark boundaries and discontinuity boundaries very well. The problem of boundary localization errors in ultrasound images was also well resolved. Our model had a finer and more robust performance.

   \subsubsection{Comparison with vascular ultrasound image segmentation models}
   In order to ensure fairness, we compared the existing vascular ultrasound image segmentation models \cite{xie2019vessel,szarski2021improved,yang2018ivus,yang2019robust}. The experimental results are shown in Table \ref{tab2}. The experiments show that our method is better than the previous methods. In terms of segmentation performance, our Dice is improved by 1.9 compared to the best existing model. In terms of computational resource usage, our model is much smaller compared to models with Dice greater than 60. Considering both together, our model has an absolute advantage. The experimental result shows that our network has better segmentation performance and is easier to apply.
  \begin{table}[]
  	\caption{Quantitative comparison of different vascular ultrasound image segmentation models, using evaluation metrics for both the model itself and the segmentation results obtained from the model.}
  	\label{tab2}
  	\begin{center}
  		\resizebox{\columnwidth}{!}{		
  			\begin{tabular}{cccccc}
  				\toprule
  				Model                     & Dice & mIoU & BIoU & Params(M) & FLOPs(G) \\ \midrule
  				M. Szarski et al. \cite{szarski2021improved} & 58.5 & 69.2 & 42.4 & 0.3       & 0.7      \\
  				J. Yang et al. \cite{yang2018ivus}    & 61.1 & 70.8 & 45.2 & 21.3      & 194.9    \\
  				J. Yang et al. \cite{yang2019robust}   & 63.5 & 72.0 & 47.6 & 8.3       & 34.7     \\
  				M. Xie et al. \cite{xie2019vessel}     & 63.7 & 72.1 & 48.0 & 46.8      & 19.3     \\ \midrule
  				BDNet (ours)                      & 65.6 & 73.2 & 50.0 & 1.3       & 1.9      \\ \bottomrule
  			\end{tabular}
  		}
  	
  	\end{center}
  \end{table}
   
   The visualization results of the model are shown in Fig. \ref{fig5}. We can find that our model has better robustness in the case of discontinuous boundaries and darker boundaries. Our model can realize the connection of discontinuous boundaries and achieve accurate segmentation of dark boundaries. The model is more resistant to noise interference and segmentation is more accurate.

   \subsection{Ablation experiment}
   In the previous section, we experimentally demonstrated the merits of our model. In this section, we will discuss the important modules of the model and the role of the loss function at the level of experimental results, and how each module affects the results.

   \subsubsection{The effect of different modules on the segmentation results}

       \begin{figure*}
    	\begin{center}				
    		\includegraphics[width=0.8\textwidth]{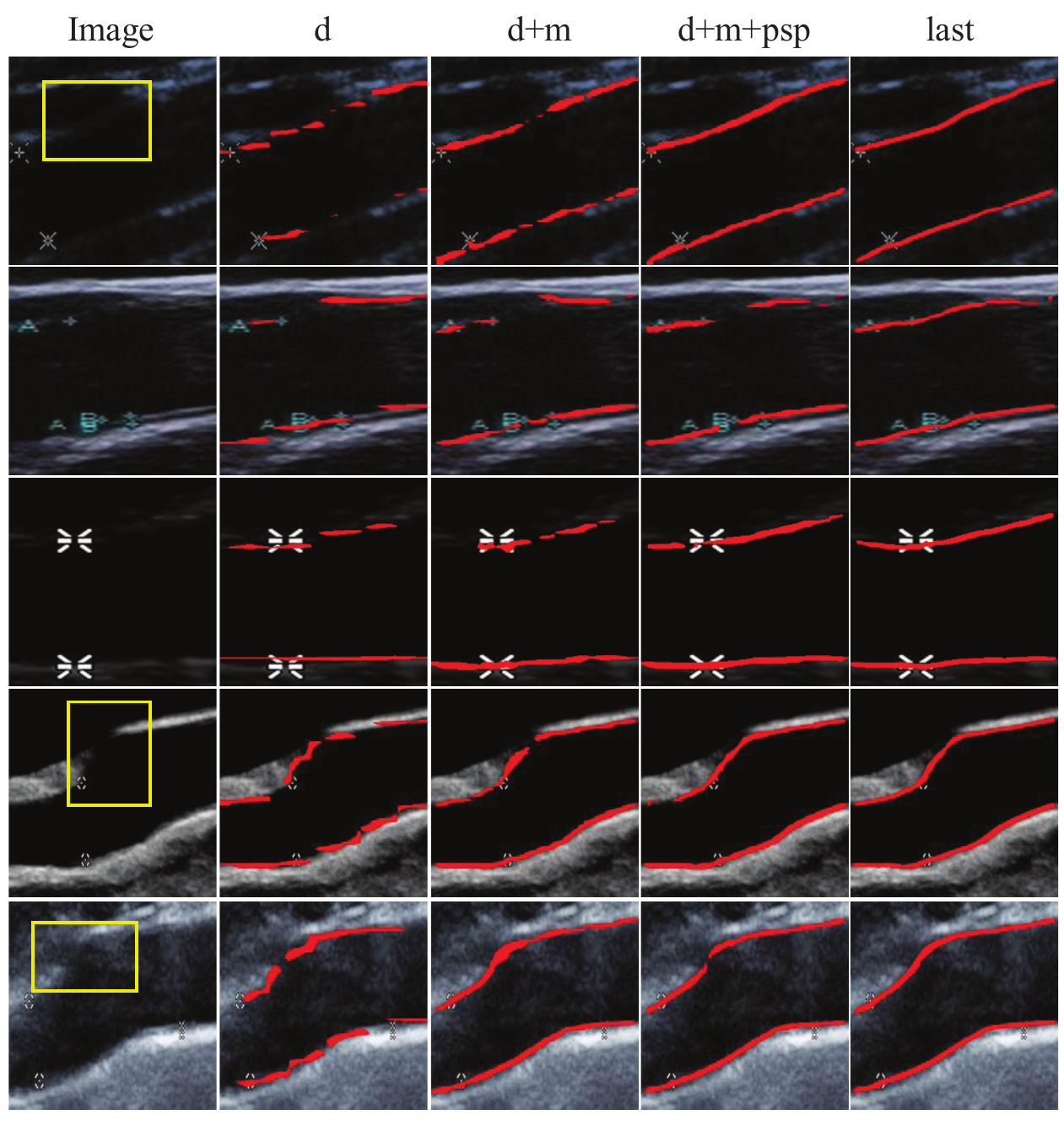}		
    		\caption{Visualization of segmentation results for models with different modules.}
    		\label{fig6}
    	\end{center}		
    \end{figure*}
      \begin{table}[]
   	\caption{Comparative experimental results of models with different module (d indicates downsampling module, m indicates multi-scale fusion module, psp indicates PSP module, and b indicates boundary refinement module).}
   	\label{tab5}	
   	\begin{center}
   		\resizebox{\columnwidth}{!}{		
   			\begin{tabular}{cccc}
   				\toprule
   				Model               & Dice        & mIoU       & BIoU        \\ \midrule
   				d                   & 44.3        & 62.6       & 28.9        \\
   				d+m                 & 62.7(+18.4) & 71.6 (+9)  & 46.5(+17.6) \\
   				d+m+psp             & 64.4 (+1.7) & 72.5(+0.9) & 48.4 (+1.9) \\
   				last(d+m+psp+b) & 65.6(+1.2)  & 73.2(+0.7) & 50.0(+1.6)  \\ \bottomrule
   			\end{tabular}
   		}
   	\end{center}	
   \end{table} 
   To demonstrate the importance of each module, we experimented with gradually adding modules. In the first experiment, we just used a simple downsampling module for training and prediction. Then, added the multi-scale fusion module. After that, the PSP module was added. Finally, the boundary refinement module was added. The experimental results are shown in Table \ref{tab5} and Fig. \ref{fig6}. In Table \ref{tab5}, we can find that the performance of our model was significantly improved after adding the multi-scale fusion module. With the addition of the PSP module and the boundary refinement module, our model had a good improvement in both. In Fig. \ref{fig6}, we can see that with the addition of the multi-scale module our model started to resolve images with dark boundaries. After adding the PSP module our model can solve the problem of boundaries discontinuously smoothly. 
  
\begin{figure}
   	\begin{center}
   		\includegraphics[width=0.9\columnwidth]{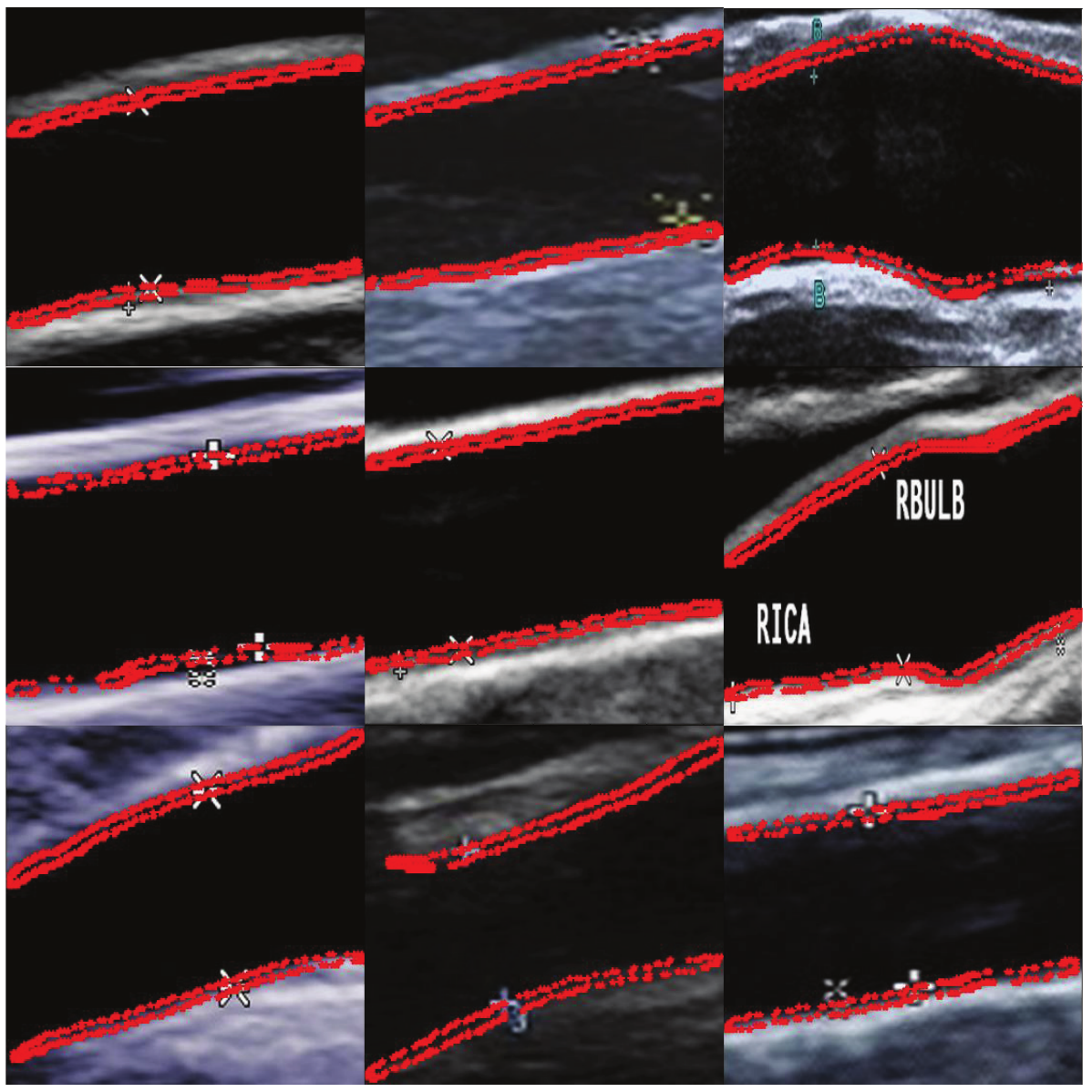}		
   		\caption{The distribution of uncertainty points in the image.}
   		\label{fig7}
   	\end{center}	
   \end{figure}

   As for the boundary refinement module, we can find that boundary refinement improved the segmentation accuracy of the model from the experimental result in Table \ref{tab5}. However, the boundary refinement module was not obvious in terms of the visualization effect. To demonstrate the usefulness of this module, we visualized the positions of the first 640 uncertain points extracted by the boundary refinement module, as shown in Fig. \ref{fig7}. We can see that the uncertain points extracted by the model are boundary points. Thus, this shows that our model is indeed able to locate the boundary points, thus preventing the boundary information from being smoothed due to upsampling and thus leading to incorrect boundary locations.
   
   \subsubsection{The effect of loss function on the segmentation results}

\begin{figure}
	\begin{center}		
		\includegraphics[width=0.9\columnwidth]{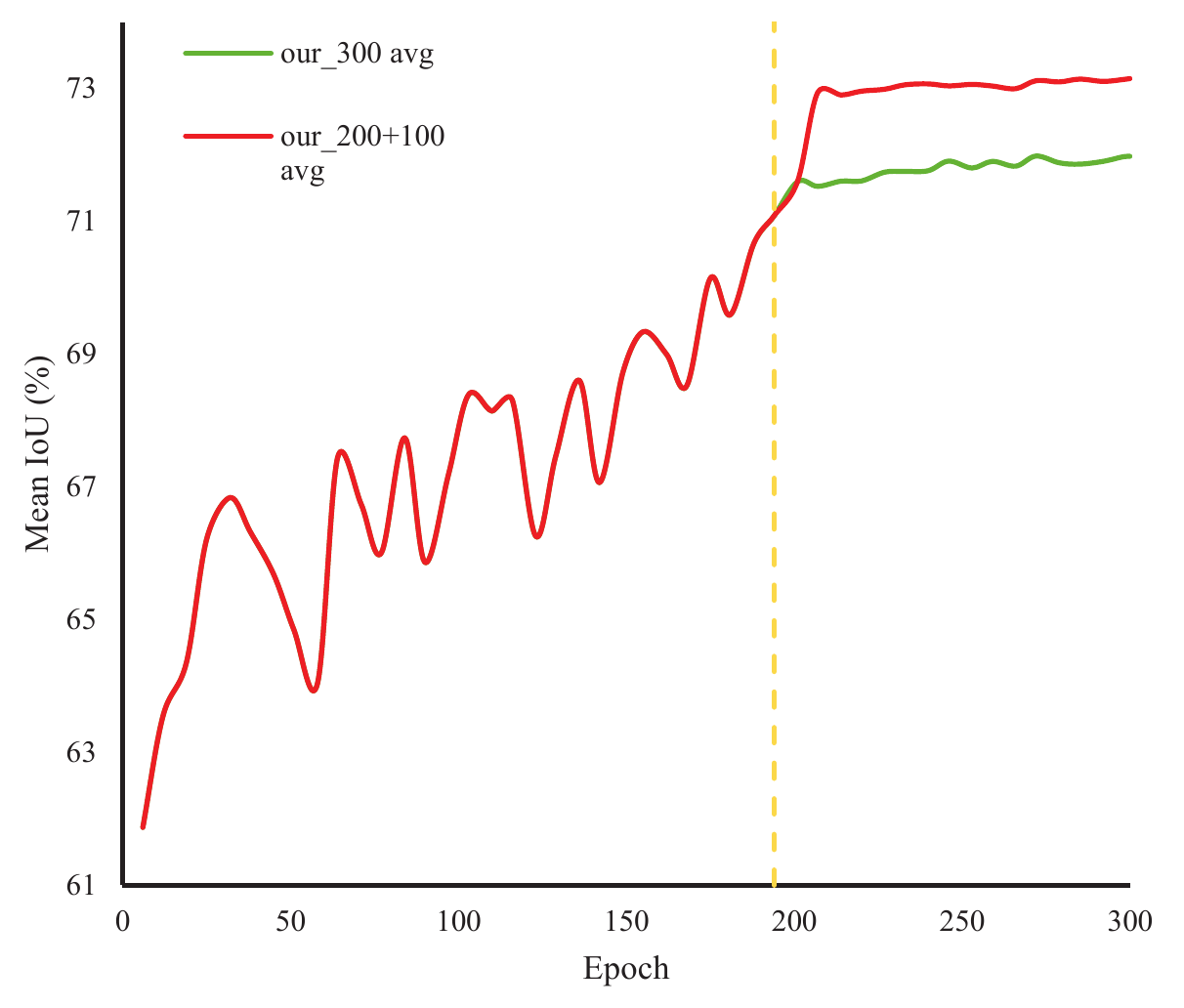}
		\caption{Variation of mIoU during the training process for models using different loss functions.}
		\label{fig8}
	\end{center}	
\end{figure}

   To test the effectiveness of our loss function, we did the following experiment. In one model, 300 epochs were trained with $L_{CE}$ and $L_{PCE}$ only. In the other model, we trained 200 epochs with $L_{CE}$ and $L_{PCE}$ and 100 epochs with $L_{CE}$ and $L_{LS}$ for optimization. The experimental results are shown in Fig. \ref{fig8} and Table \ref{tab6}.

         \begin{table}
	\caption{Quantitative comparison of the segmentation performance of the model with different loss functions. Our\_300 means just 300 epochs trained with $L_{CE}$ and $L_{PCE}$. Our\_200+100 means 200 epochs trained with $L_{CE}$ and $L_{PCE}$, and then 100 epochs trained with $L_{CE}$ and $L_{LS}$.}
	\label{tab6}	
	\begin{center}
		\resizebox{\columnwidth}{!}{		
			\begin{tabular}{cccc}
				\toprule
				Model        & Dice       & mIoU       & BIoU       \\\midrule
				Our\_300     & 62.6       & 72.0       & 47.0       \\\midrule
				Our\_200+100 & 65.6(+3.0) & 73.2(+1.2) & 50.0(+3.0) \\ \bottomrule
			\end{tabular}
		}
	\end{center}	
\end{table}
  
   As shown in Fig. \ref{fig8} and Table \ref{tab6}, after 200 epochs, the accuracy of the model with the addition of $L_{LS}$ increased rapidly and remained stable. After the training was completed, the Dice of the model with $L_{LS}$ improved by 3.0, mIoU improved by 1.2, and BIoU improved by 3.0. Thus, the combination of our loss functions is effective.
   
    \subsubsection{The effect of lightweight strategy on model parameters and segmentation results}

   To reduce the number of parameters in the model and to avoid excessive hardware requirements, we designed a lightweight downsampling module. In this section, we will see how our design worked through experimental results. In the comparison experiments, we replaced the depthwise separable convolution in the model with the conventional convolution. The other parts remained unchanged. The experimental results are shown in Table \ref{tab3}. $M_{dsconv}$ is the model we designed. $M_{conv}$ is the model in which the depthwise separable convolution is replaced by the conventional convolution. $All$ denotes the whole model. $P_{d}$ denotes the proportion of the downsampling module in the whole model.
   \begin{table}[]
   	\caption{Analysis of the number of parameters for each module of our model and the traditional convolution-based model.}
   	\label{tab3}	
   	\begin{center}
   		\resizebox{\columnwidth}{!}{		
   			\begin{tabular}{ccccccc}
   				\toprule
   				& Model   & downsampling & fusion & boundary refinement & $All$           & $P_{d}$             \\ \midrule
   				\multirow{2}{*}{FLOPs(G)}  & $M_{dsconv}$ & 0.19     & 0.98           & 0.74      & $\bm{1.90}$ & $\bm{9.72}$ \\
   				& $M_{conv}$   & 2.43     & 0.98           & 0.74      & 4.15          & 58.59          \\ \midrule
   				\multirow{2}{*}{Params(M)} & $M_{dsconv}$ & 0.50     & 0.64           & 0.19      & $\bm{1.32}$ & $\bm{37.55}$ \\
   				& $M_{conv}$   & 11.24    & 0.64           & 0.19      & 12.07         & 93.15          \\ \bottomrule
   			\end{tabular}
   		}
   		\vspace{-2em}
   	\end{center}	
   \end{table}
   
   The experimental results show that the downsampling module of our model has very little weight in the whole model. We can design more special structures that fit the model's function and thus increase the model's performance while ensuring a small number of parameters. In the traditional method, the downsampling module takes up too much weight in the model. If there is a limit on the number of parameters, it is difficult to have more space for other designs.

   \begin{table}[]
   	\caption{Segmentation results of our model and traditional convolution-based models.}
   	\label{tab4}	
   	\begin{center}
   		\resizebox{\columnwidth}{!}{		
   			\begin{tabular}{cccccc}
   				\toprule
   				Model   & Dice       & mIoU       & BIoU       & Params & FLOPs \\\midrule
   				$M_{conv}$   & 65.8       & 73.2       & 50.2       & 12.1   & 4.2   \\
   				$M_{dsconv}$ & 65.6(-0.2) & 73.2(-0.0) & 50.0(-0.2) & $\bm{1.3}$    & $\bm{1.9}$ \\ \bottomrule
   			\end{tabular}
   		}
   		\vspace{-2em}
   	\end{center}	
   \end{table}

   After the above results, we can find that our model achieved lightweight. In the following, we experimentally observed the decrease in the number of parameters affects the segmentation performance. Table \ref{tab4} shows the experimental results of the segmentation performance of the two models. We find that our model has a slight decrease in performance compared to $M_{conv}$. However, the number of $M_{conv}$'s parameters is 9 times higher than that of our model. All things considered, it is worthwhile to sacrifice a little performance and significantly reduce the number of parameters. It proves that our lightweight is meaningful.
   
   \section{Discussion}
   \label{Section IV}
   In this study, we proposed four modules for vascular wall segmentation. The first one is that we designed a feature extraction module to extract multi-scale features and fused them, generated different receptive field features and fused them, and introduced more detailed and contextual information. After that, we used the boundary refinement module to prevent the effect of upsampling on the boundary position. Then, we designed a loss function to avoid the effect of class imbalance on the model and focused the optimization of the model on vascular wall segmentation. Finally, we designed a lightweight downsampling module to reduce the parameters of the model.
   
   Compared with other models, our model has advantages in terms of the number of parameters and segmentation effects, as shown in Table \ref{tab1} and Table \ref{tab2}. The Dice of our model is 2.3 higher compared to the best generic model and 1.9 higher compared to the best vascular ultrasound image segmentation model. Our model has only one-twentieth or less of the number of parameters of these two models. We think that there are two reasons for this situation. One is that we have made some lightweight designs for the model. The other is our special design corresponding to the characteristics of ultrasound vascular images. 
 
   We had used depthwise separable convolution instead of traditional convolution for downsampling. This approach can greatly reduce the number of parameters of the model. As shown in Table \ref{tab3}, the number of parameters decreases to one-ninth of the original one. The depthwise separable convolution has little impact on segmentation performance, as shown in Table \ref{tab4}.
 
   For the problem of dark boundaries, Xie et al. \cite{xie2019vessel} extracted more image features by setting two encoders. Two encoders do extract features very well, but the drawback is also obvious. The number of parameters becomes correspondingly large. As is shown in Table \ref{tab2}, the number of parameters is 36 times larger than our model. This is not hardware friendly. We just used an encoder and chose to use multi-scale information fusion processing. The model combined detailed information and high-level features for prediction. With the problem of dark boundaries solved, the performance of the model is naturally improved to a larger extent. For the problem of discontinuity boundaries of the vascular wall, we extracted information from both global and local views by expanding the receptive field. Yang et al. \cite{yang2018ivus} expanded the receptive field by changing the convolution kernel size to 5×5. The method has some effect, but the expansion of the field of view is limited. Our model combined global and local information from multiple fields of view for feature fusion and extraction. Therefore, the results are better. Most importantly, we focused on the boundary positioning error problem in the model design. Accurate boundary locations are obtained by correcting the prediction results of the upsampled boundary points. In addition, the boundary point predictions were added to our loss function to ensure that the model can predict the boundary points well. Finally, for the problem of predicting boundary class imbalance, we used the lovász-softmax loss to solve it. The mIoU principle was used to prevent interference from the background.
  
   One more point is worth emphasizing. Many of the techniques in our solution are not limited to the processing of vascular ultrasound images. For example, the optimization of parametric quantities applies to many image processing application problems. Many image processing models encounter the problem of incorrect boundary locations. We hope that our model can be an inspiration to other researchers.
  
   But our model also has some limitations that we need to improve. Our model is segmented based on image information now. There is still little information in a single image. The points output by the boundary refinement module is generated based on the boundary information of the images. The structure of blood vessels is regular. We can incorporate the physician's a priori knowledge into our model. Afterward, we will consider analyzing the positional relationships between these generated points. The geometry of the vessels will be fitted according to the positional relationships to reconstruct our vascular walls. We will design a segmentation model with better robustness.
   
   \section{Conclusions}
   \label{Section V}
   In this paper, our goal is to design a boundary-delineation network (BDNet) to segment the vascular walls of ultrasound vascular images. We solved the incorrect segmentation of low-quality images with dark boundaries and discontinuous boundaries by mixing image features of multi-scale and different receptive fields. We used the boundary refinement module to precisely locate the boundary points and re-predict the boundaries to obtain the correct boundary locations. We optimized the model by loss function to make the segmentation smoother and more accurate. Finally, we optimized the model with parameters to make it lightweight and application-friendly. After the above processing, our model has good robustness and a finer segmentation effect. Through experimental comparisons of the dataset, our method achieved state-of-the-art segmentation performance. Based on the current model results, our model is more applicable to the segmentation of object boundaries. Therefore, it will consider trying in other cases of boundary segmentation of medical images in the future.

			\end{document}